\documentclass[manuscript]{aastex}

\usepackage[usenames]{color}
\usepackage{epsfig,graphicx}

\newcommand{\rsun}{R$_{\odot}$}

\newcommand{\lya}{Ly~$\alpha$}

\newcommand{\hi}{H~{\sc{i}}}
\newcommand{\ovi}{O~{\sc{vi}}}
\newcommand{\mgx}{Mg~{\sc{x}}}

\shorttitle{Observations and models of \mgx\ ions}
\shortauthors{Ofman, Abbo, Giordano}

\begin{document}

\title{Observations and models of slow solar wind with Mg$^{9+}$ ions in quiescent streamers}
\author{L. Ofman\altaffilmark{1,2,3}, L. Abbo\altaffilmark{4}, S. Giordano\altaffilmark{4}}
\altaffiltext{1}{Catholic University of America, Washington, DC
20064} \altaffiltext{2}{NASA Goddard Space Flight Center, Code 671,
Greenbelt, MD 20771} \altaffiltext{3}{Visiting Associate Professor, Department of
Geophysics and Planetary Sciences, Tel Aviv University, Tel Aviv 69978,
Israel}
\altaffiltext{4}{INAF, Astrophysical
Observatory of Turin, Italy}

\begin{abstract}
Quiescent streamers are characterized by a peculiar UV signature as pointed out by the results from the observations of the Ultraviolet and Coronograph Spectrometer (UVCS) on board SOHO: the intensity of heavy ion emission lines (such as \ovi) show dimmer core relative to the edges. Previous models show that the structure of the heavy ion streamer emission relates to the acceleration regions of the slow solar wind at streamer legs and to gravitational settling processes in the streamer core. 
Observations of Mg$^{9+}$ ion EUV emission in coronal streamers at solar minimum were first reported by the UVCS instrument. The \mgx\ 625\AA\ emission is an order of magnitude smaller than the \ovi\ 1032\,\AA\ emission, requiring longer exposures to obtain statistically significant results. Here, \mgx\ coronal observations are analyzed and compared, for the first time, with the solar minimum streamer structure in hydrogen and \ovi\ emissions. We employ the 2.5D three-fluid model, developed previously to study  the properties of O$^{5+}$ ions in streamers, and calculate for the first time the density, temperature, and outflow structure of Mg$^{9+}$ ions in the solar minimum streamer. The Mg$^{9+}$ ions are heated by an empirical radial heating function constrained by observations of the kinetic ion temperature obtained from \mgx\ emission line profiles. The detailed structure of  Mg$^{9+}$ density, temperature, and outflow speed is determined by the Coulomb momentum and energy exchange as well as electromagnetic interactions with electrons and protons in the three-fluid model of the streamer. The results of the model are in good qualitative agreement with observations, and provide insights on the possible link between the magnetic structure of the streamer,  slow solar wind sources, and relative abundances of heavy ions.
\end{abstract}
\section{Introduction}

One of the main results obtained from the observations performed by the Ultraviolet and Coronograph Spectrometer (UVCS) on board SOHO was the dramatic difference in the line profiles and the emission variation in latitude across quiescent streamers of heavy ions such as \ovi\  1032\,\AA\  compared to the \hi\,\lya\,1216\,\AA\ emission \citep[e.g.,][]{Koh97,Ray97,Noc97,Koh99,Mar01,Zan01,Str02,Uzz03,Uzz04,Vas05,Uzz06,Uzz07,Aki07}. This variation in emission was attributed to the variation in heavy ion relative abundance across streamers and to slow solar wind acceleration regions in several numerical multi-fluid models \citep{Ofm00a,Ofm04a,CL04,Mor08,OAG11}. Note, that modeling the self consistent interaction between heavy ions, protons, and electrons in coronal streamers requires multi-fluid models as the next level of physical approximation beyond a single-fluid MHD. In these models the electrons, protons, and heavy ions are described as electromagnetically, and collisionally interacting charged fluids with overall quasi-neutrality. The electron inertia is usually neglected compared to the proton inertia, and the electron momentum equation is used to solve for the electric field (generalized Ohm's law). Separate coupled continuity, momentum, and energy equations are solved for each particle species together with Maxwell's equations. 

Recently, \citet{OAG11} used the 2.5D (azimuthally symmetric) 3-fluid model to study the UV emission from H I
\lya\,1216\,\AA\ and \ovi\  1032\,\AA\ spectral lines in a streamer belt during solar minimum (CR 1913) observed with UVCS/SOHO.  Dipole magnetic field configuration was included in the initial state and preferential heating
for O$^{5+}$ ions was considered with coupled polytropic energy equations for electrons protons and ions. The model relaxed to a steady state with a coronal streamer belt and a slow solar wind that included the heavy ions. The results of the model provide information on the slow solar wind acceleration regions in the streamer belt, the interactions between the species, and the velocity, density, and temperature variations of the electrons, protons, and O$^{5+}$ ions in the streamers. The model results were used to synthesize UV observables in the above spectral lines and very good agreement was found with UVCS observations. \citet{Abb10a} used the results of single-fluid MHD mode to determine the edges of streamers and compared the results with the UVCS observational of \ovi\ emission variation across a streamer. First detailed  comparisons between \ovi\ emission in streamers and 3-fluid model results were done by \citet{Ofm00a,Ofm04a}. Recently, a preliminary comparison to multi-fluid model was performed by \citet{Abb10b} that was extended in a detailed study by \citet{OAG11}. Since the interactions between the various heavy ion species as well as their effect on protons and electrons is negligible due to their low relative abundances, one has the  freedom to choose any single minor heavy ion species in the three-fluid model calculations. Here, we extend our previous studies by introducing  Mg$^{9+}$ as the heavy ion species, and investigate the details of the ion streamer density, temperature, velocity, and UV emission. This is the first detailed observational study of streamer emission and images in \mgx\ 625\,\AA, and the first detailed 2.5D three-fluid modeling of streamers that include Mg$^{9+}$  ions in addition to protons and electrons.
In particular, the aim of this study is to compare the observed radial and latitudinal emission profiles with the expected ones
from multi--fluid MHD model of coronal streamer as seen in \hi\,\lya\, and in \mgx\,to validate the model results and better understand the origin of the slow solar wind.

\section{Quiescent Streamers Observations in \mgx\ 625\,\AA}

The doublet of UV coronal emission lines from Mg$^{9+}$ at 609.76\,\AA\, and 624.93\,\AA, 
here called \mgx\,  610\,\AA\ and \mgx\, 625\,\AA\ can be detected by UVCS spectra--coronagraph at the 
second order in the channel optimized for \ovi\, lines at 1032\,\AA\, and 1037\,\AA, that includes
longer wavelength from its redundant optical path. The expected most intense \mgx\, 610\,\AA\ emission 
falls on the red wing of the very bright and wide \hi\ \lya\, at 1215.67\,\AA, therefore, 
the spectral analysis is more difficult due to lower signal to noise ratio than for  \mgx\, 625\,\AA\ which falls
about 30\,\AA\ away from the \hi\  line. Thus, we focus on \mgx\, 625\,\AA\ emission and in particular on the study of the quiescent coronal streamers
that show the typical intensity dimming in the central region of the streamers as observed in \ovi\, lines, in order
to answer the question whether this dimming is present in emission lines from heavier ions, such as \mgx. Multi-fluid models have shown that the density structure of the heavy ions are closely related to the structure of the slow solar wind in streamers \citep[e.g.,][]{Ofm00a,Ofm04a}.  The quiescent streamers are well defined at the minimum of solar activity at the equatorial latitudes,
in particular at the minimum of cycle 22 (1996--1997) the streamer belt was very stable and could be modeled reasonably well by 2.5D model with initial dipole magnetic field \citep[e.g.,][]{OK10,OAG11}.

Because of the low Magnesium abundance in the solar corona ($\simeq4\times 10^{-5}$ relative to Hydrogen for all ionization states) and the reduced 
second order lines instrument efficiency, the photon count rates from \mgx\ lines are low, therefore to reach a good statistics, a very long exposure time is required. Alternatively, multiple observations of streamers with shorter exposures are combined to reconstruct an 
averaged \mgx\, 625\,\AA\ emission radial and latitudinal profiles.

Almost daily observations of \mgx\, 625\,\AA\, line were performed with UVCS from 1997 July 31 to November 25.
In that period we identified the long--lived quiescent streamers by looking 
at LASCO C2 observations of west and east limbs. Next, we analyzed the corresponding intensity profiles of \hi\ \lya\, and \ovi\, 1032\,\AA\ in order to identify the cases of evident core--dimming in \ovi.
Thus, for these streamers, we consider the \mgx\, 625\,\AA\ spectra to determine the
intensity profiles.

\begin{figure}
\center
\includegraphics[width=16cm]{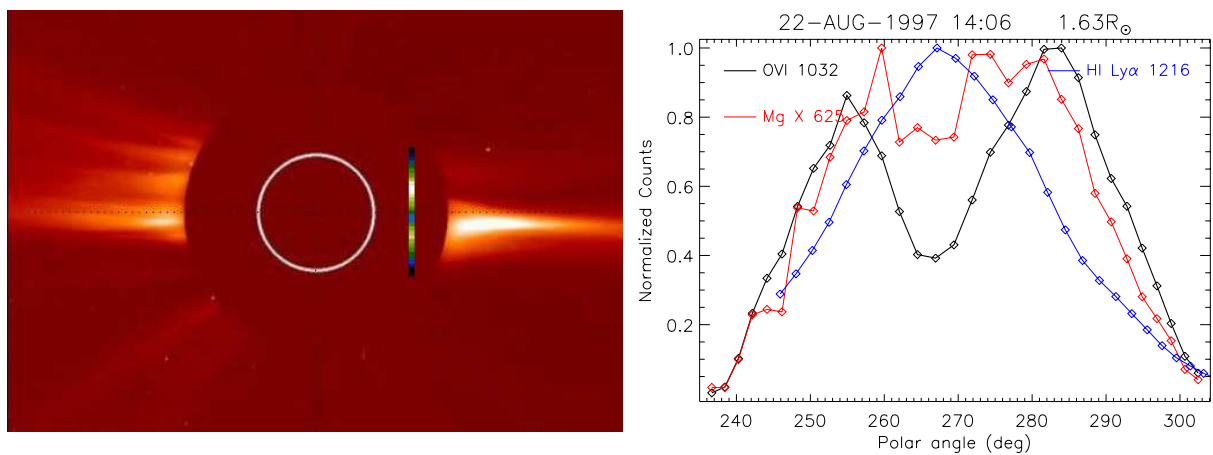}
\caption{Quiescent streamer observation at solar minimum in visible light and UV performed on August 22, 1997. Left panel: White light image from LASCO C2 with superimposed the O VI 1032\AA\ intensity image along UVCS slit. Right panel: Intensity profiles along the instrumental slit at 1.63 R$_\odot$ of \hi\,\lya\  (blue), \ovi\,1032\AA\ (black) and \mgx\,625\AA\ (red) spectral lines.}
\label{uvcs_strm:fig}
\end{figure}

A sample of the streamer observations considered in the study is shown in the left panel of Figure~\ref{uvcs_strm:fig}, 
where the background image is the visible light scattered by coronal electrons from LASCO C2 observations,
and the vertical strip is the \ovi\ 1032\,\AA\ line intensity detected by UVCS. The data were  obtained on August 22, 1997.
The right panel of Figure~\ref{uvcs_strm:fig} shows the latitudinal intensity profiles at 1.63 \rsun\,of \hi\,\lya\, (blue), \ovi\, 1032\,\AA\ (black) and \mgx\, 625\,\AA\ (red) spectral lines. The depletion in \ovi\  emission at the center of the streamer is evident and there is indication of the core--dimming also in \mgx\, emission line, although of lower magnitude.

The \mgx\, 625\,\AA\ line intensity profile as a function of the radial distance has been derived using  UVCS data from August 23 and September 26, 1997 that were obtained  
up to heliocentric distances of 2.75 and 2.28 \rsun, respectively. In the latter case we were able to reconstruct the \mgx\, coronal image
from a streamer scan that clearly shows the presence of a core--dimming from Mg$^{9+}$ ion emission. The \mgx\ core--dimming is less pronounces in comparison with the dimming seen in O$^{5+}$ ion emission (Figure~\ref{MgxUV:fig}).
We determine the width
of \mgx\, spectral line by analyzing observations with narrow slits. 
There are only few UVCS streamer observations of \mgx\ 625\,\AA\ line with narrow slit during the solar minimum;
moreover, these observations do not provide good spatial resolution data, because of the narrow slit 
and resulting low count rate. Therefore, it is difficult to determine the spectral parameters  in streamer
core and legs, separately. 
The values averaged over about 500$\arcsec$ around the streamer axis are reported
in Table~1 where the line widths are given in terms of most probable velocity ($V_{1/e}$) and of kinetic temperature, $T_{k,i}$, both for \mgx\, and \ovi\, lines. Note that the relation between  $T_{k,i}$ and $V_{1/e}$ is given by 
 $T_{k,i}=\frac{A_i m_p}{2 k_B} V_{1/e}^{2}$, where $A_i$ is the atomic mass number of the ions, $m_p$ is the proton mass, and $k_B$ is Boltzmann's constant.

In quiescent streamer at solar minimum we find that $V_{1/e}$ of Mg$^{9+}$ ions is comparable or slightly larger than that 
of O$^{5+}$ ions, that is $V_{1/e,MgX} \geq V_{1/e,OVI}$ and the kinetic temperature of
Mg$^{9+}$ ions is much larger than that of O$^{5+}$ ions. While the mass ratio of Mg$^{9+}$ ions and O$^{5+}$ ions is $1.5$, the kinetic temperature ratio is on average about a factor of two (i.e., $T_{k,MgX}/T_{k,OVI}\sim 2$, as evident in Table~1). The high ion temperatures suggest more than mass proportional heating, in agreement with previous studies in coronal holes \citep[e.g.,][]{Koh99,CFK99}. However, the temperature ratio between the two ions in streamers has the opposite trend compared to the observations in coronal holes that find higher O$^{5+}$ temperatures than Mg$^{9+}$ temperatures \citep[e.g.][]{Koh99,Ess99}.

\section{Three-fluid model: boundary conditions and numerical results}
We employ the 2.5D three-fluid model used recently to model coronal streamers with O$^{5+}$ ions at solar minimum \citep[see,][for further details of the model]{OAG11}, with brief description of the model provided below. The streamer calculation is initialized with a potential dipole magnetic field normalized by $B_0=7$ G, initial temperature $T_0=1.6 MK$, and normalization density $n_0=5\times 10^8$ cm$^{-3}$, and the corresponding Alfv\'{e}n speed $V_A=683$ km s$^{-1}$. These values are used to calculate the solar wind velocity with the isothermal Parker's solution as the initial state for the three-fluid model. A polytropic energy equation with $\gamma=1.05$ is solved  for each particle species with the addition of collisional energy exchange terms, and with observationally constrained empirical heating term for Mg$^{9+}$ ions to model their preferential heating (see below). The mass continuity and momentum equations are solved for protons and heavy ions, while charged neutrality is used to obtain the electron density. The neglect of electron inertia is used obtaining the generalized Ohm's law, and the electric field is used in the momentum equations that include Coulomb friction terms between the various species. The electron velocity is used in the induction equation and the gyro-motions of the ion fluids are neglected in the low-frequency (MHD) limit. The boundary conditions are open at the outer boundary, and fixed field at the lower boundary, with extrapolated density, temperature, and velocity (based on the method described in \citet{SN76}). The resolution is 320 grid cells in $\theta$ in the (approximate) range 0 to $\pi$, and 1028 in $r$ in the range $(1,\,8)$ \rsun.

We modify the model to study the dynamics of Mg$^{9+}$ ions by adopting the parameters for the Mg$^{9+}$ ions as the third fluid; i.e., the charge $Z=9$ and atomic mass to $A=24$, and we use the relative abundance of $n_{Mg9+,0}=5\times10^{-6}$ \citep{Maz98} at $r=1$\rsun\ of the model. Note, that as in previous three-fluid models we assume that the ionization state remains fixed, i.e., the ionization and recombination of Mg$^{9+}$ is at steady state, even though the electron temperature is not constant. This is justified by the result that at steady state the closed field region of a quiescent streamer is in hydrostatic equilibrium, with (nearly) constant electron temperature \citep[e.g.,][]{PK71}, while in open field region the ionization state of Mg$^{9+}$ ions becomes frozen-in within 1.5$R_\odot$ when ion lifetime is compared to the solar wind expansion time \citep[e.g.,][]{Wit82,Ess99}. The initial Mg$^{9+}$ relative abundance is uniform in the computational domain. The initial Mg$^{9+}$ temperature $T_{i,0}=42$ MK, and the initial electron and proton temperatures $T_{e,0}=T_{p,0}=1.6$ MK are uniform.  Note, that $T_{i,0}$ is taken to be large so that the initial hydrostatic Mg$^{9+}$ ion density does not decreases to zero at the top of the streamer. However, as the solution evolves to steady state, $T_i$ decreases rapidly and the density structure coupled to the protons and electrons becomes more realistic. We have iterated  on the value of $T_{i,0}$ to provide the best match with the observed \mgx\ emission in the streamer. In addition, we study the parameters of the ion empirical heating function given by 
\begin{eqnarray}
&& S_i=S_{i0}(r-1)
e^{-(r-1)/\lambda_{i0}}, 
\label{Mg9heat:eq}
\end{eqnarray} 
where the parameters $S_{i0}$ and $\lambda_{i0}$ determine the magnitude and the decrease of the heating rate with distance, $r$. The heating function parameters are adjusted to match the $T_{k,MgX}$ values in the center of the streamer at the steady state determined from the line profiles given in Table~\ref{tk:table}. We found that $S_{i0}\approx7.5\times10^2$ and $\lambda_{i0}=0.225$\rsun\ provide the best match with observations for Mg$^{9+}$ ion temperature structure in the present model. It is interesting to note that in  \citet{OAG11} the value of $S_{i0}$ for O$^{5+}$ ions was $\sim$26 times smaller, and the value of $\lambda_{i0}$ was a factor of two larger than in the present study. The three-fluid model is evolved until the streamer solution reaches a (quasi-) steady state, with the typical closed field and open field region structures separated by a current sheet. We use the term quasi-steady state since after the rapid evolution stage the streamer still evolves slowly at this stage. The Mg$^{9+}$  temperature decreases quickly to much lower values while the electron and proton temperatures do not strongly vary as the solution reaches the steady state. Note, that the model  is aimed at studying the slow solar wind in the streamer, and does not include the additional momentum input in open field regions required for fast solar wind solutions.

\begin{figure}
\center
\includegraphics[width=15cm]{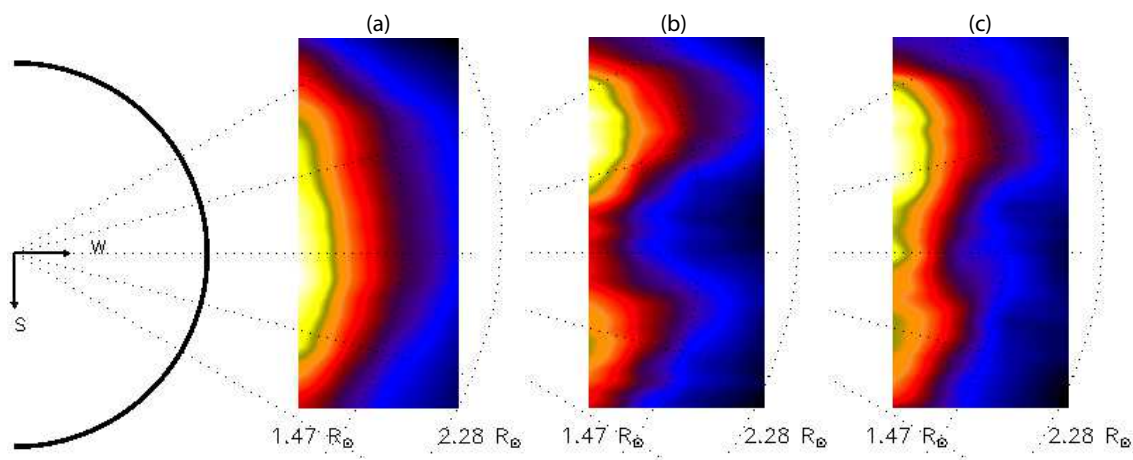}
\caption{Streamer intensity images from 1.47 to 2.28 \rsun, as observed by UVCS on Sep. 26, 1997, in (a) \hi\,\lya\, (b) \ovi\, 1032\AA, and (c) \mgx\, 625\AA. The left panel shows the corresponding solar coordinates.}
\label{MgxUV:fig}
\end{figure}

\begin{figure}
\center
\includegraphics[width=6in]{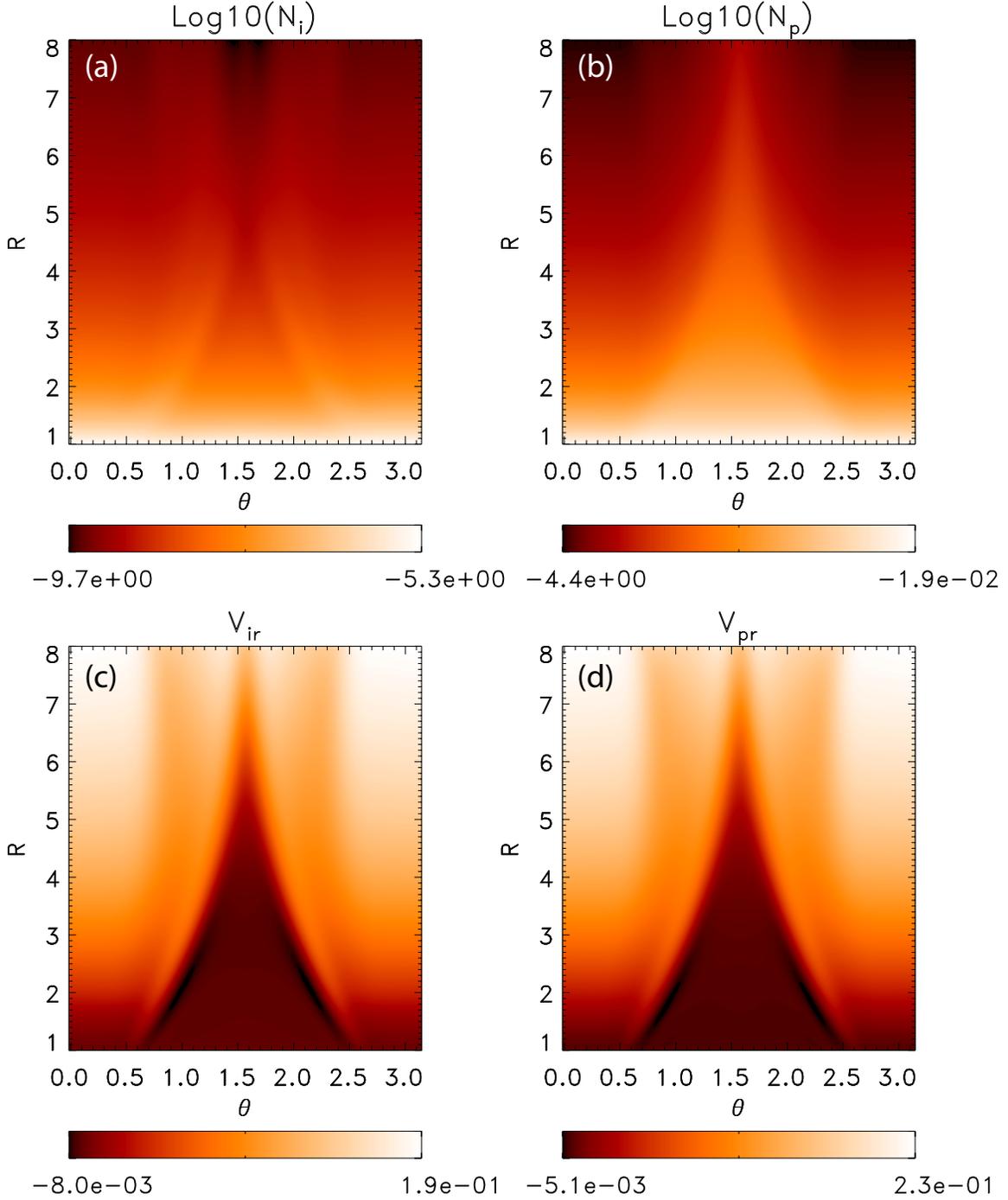}
\caption{The density of (a) protons, and (b) Mg$^{9+}$ ions in the streamer calculated with the three-fluid model at $t=98.1 \tau_A$. The corresponding radial velocity $V_r$ in units of $V_A$ of (c) protons, and (d) Mg$^{9+}$ ions. The animation of the Mg$^{9+}$ density is available in the electronic version of this article.}
\label{nVr_Mg9:fig}
\end{figure}

In Figure~\ref{nVr_Mg9:fig} the densities of protons and Mg$^{9+}$ ions and the corresponding radial outflow velocities of the streamer in steady state at $t=98.1 \tau_A$ (about 30 hours) are shown. The proton density shows an increase in the closed field region where the plasma is confined and compressed by the Lorentz force produced by the current-sheet that separates the closed and open field region (i.e., pressure balance, see~\citet{PK71}). However, the Mg$^{9+}$ ion density shows the typical darkening in the closed field region of the streamer, consistent with previous model and observations of O$^{5+}$ ions, and consistent with the observed Mg$^{9+}$ emission shown in Figure~\ref{MgxUV:fig}c.  The model shows (examining the quantitative effects of the various terms) that this structure of Mg$^{9+}$ is the result of gravitational settling in the closed field region of the heavy Mg$^{9+}$ ions, and the coupling between the outflowing protons and Mg$^{9+}$ ions due to the Coulomb friction in the open field region, resulting in increase of the relative  Mg$^{9+}$ ion abundance compared to the abundance in the closed field region. The gravitational settling is limited by the quasi-neutrality of the plasma that increases the effective scale-height of Mg$^{9+}$ ions, consistent with previous results \citep{Len04,OAG11}. Temporal evolution of Mg$^{9+}$ density structure and the formation of the density dip in the core of the streamers are evident from the enclosed animation of the  Mg$^{9+}$ ion density. However, the details of the evolution and the time scale are the results of initial conditions of the idealized model, and we should expect only qualitative agreement of the evolution with observations.

\begin{figure}
\center
\includegraphics[width=6.5in]{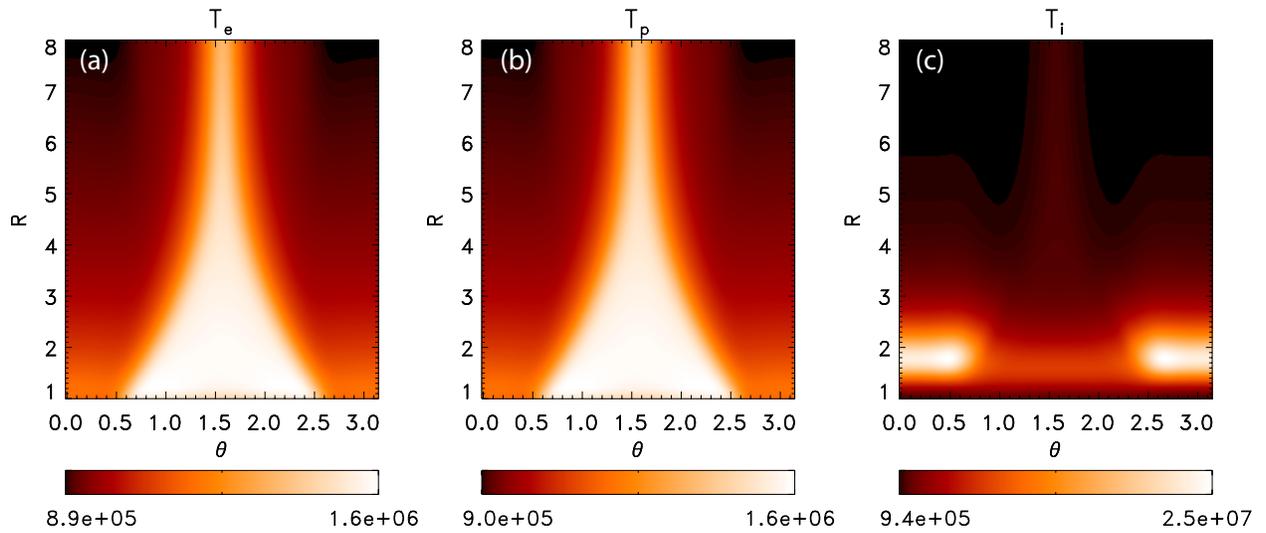}
\caption{The kinetic temperature maps of (a) electrons , (b) protons, and (c) Mg$^{9+}$ ions in the streamer calculated with the three-fluid model at $t=98.1\tau_A$.}
\label{TMg9:fig}
\end{figure}

The kinetic temperatures of electrons, protons, and Mg$^{9+}$ ions calculated with the three-fluid model at $t=98.1\tau_A$ are shown in Figure~\ref{TMg9:fig}. It is evident that the electron and proton temperature structure is similar, peaking in the closed field region of the streamer at $T=1.6MK$, consistent with observations of a coronal streamer at solar minimum and with previous three-fluid model \citep[e.g.][]{OAG11}. The electron and proton temperatures are nearly constant in the closed field regions . This is the result of hydrostatic equilibrium in forming at the steady  state of the quiescent streamer \citep[see, e.g.][]{PK71}. There is only small variation of $T_e$ and $T_p$ due to the confinement pressure exerted by the Lorentz force from the current sheet surrounding the streamer.  The nearly steady $T_e$ and $T_p$ in the core of quiescent streamer is in agreement with observations \citep[e.g.][]{Str02,Ant05,Uzz07}.

However, the temperature structure of Mg$^{9+}$ ions show very different structure. The temperature structure is the result of the application of the simple spherically symmetric, radially dependent heating function (Equation~\ref{Mg9heat:eq}) in the three-fluid model, combined with the effects of the magnetic structure of the streamer, and the collisional (Coulomb) coupling between the ions, protons, and electrons. In particular, in the closed field regions the collisional thermal energy exchange between the heated Mg$^{9+}$ ions and the cooler protons and electrons leads to lower temperature of the  Mg$^{9+}$ ions compared to their temperature outside the closed field region of the streamer. This temperature structure of the Mg$^{9+}$ ions is consistent with UVCS observations and model results for O$^{5+}$ ions \citep[e.g.,][]{OAG11}. Thus, although the simple heating function provides only radial dependence of the heating, the complex coupling between the three-fluids, the magnetic field, and the outflows leads to the structure of the latitudinal dependence of the Mg$^{9+}$ ion temperature.

\begin{figure}
\center
\includegraphics[width=5in]{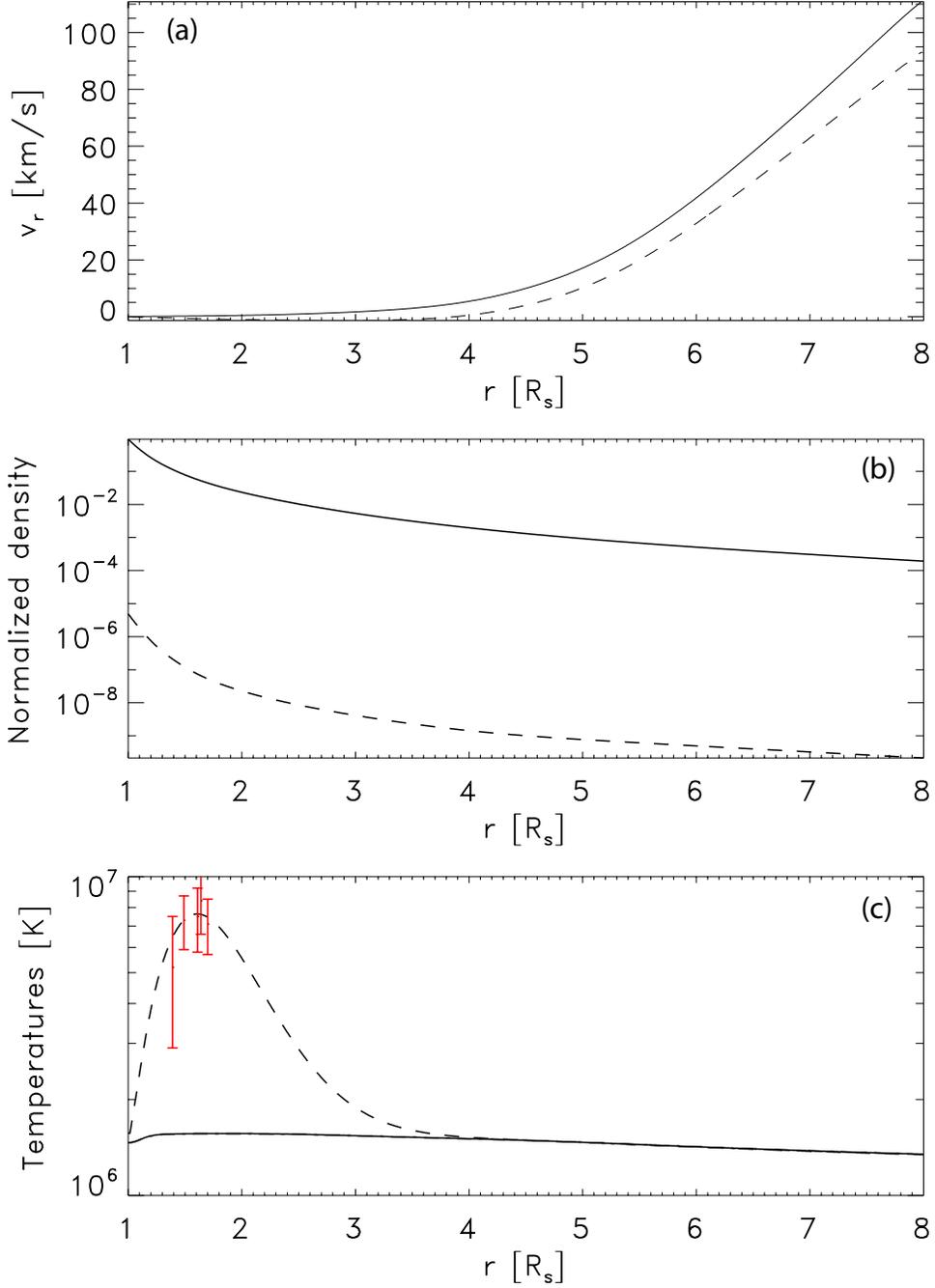}
\caption{The cross-section in the radial direction at the center of the streamer calculated with the three-fluid model at $t=98.1\tau_A$ (protons - solid; Mg$^{9+}$ ions - dashes). (a) The radial velocity. (b) The normalized density. (c) The temperature (electrons shown with dotted curve that overlaps the proton temperature). The red symbols with error bars show the kinetic temperatures of Mg$^{9+}$ given in Table~\ref{tk:table}.}
\label{vrtcr:fig}
\end{figure}

The details of the outflow speed, density, and temperature in a radial cut through the center of the solar minimum streamer are shown in Figure~\ref{vrtcr:fig}. In Figure~\ref{vrtcr:fig}a it is evident that the radial outflow speed is nearly zero in the closed field region of the streamer ($r<4R_\odot$) consistent with previous observations of O$^{5+}$ ions \citep[e.g.,][]{Str02}. In the stalk of the streamer the outflow speed of protons and Mg$^{9+}$ ions is similar, reaching $\sim100$ km s$^{-1}$ at $r=8R_\odot$. The normalized density of both, protons and Mg$^{9+}$ ions decreases rapidly with height, as expected in the solar wind streamer, shown in Figure~\ref{vrtcr:fig}b. It is interesting to note, that the Mg$^{9+}$ ion density shows clear transition between the closed field region, with rapid decrease of their density with height, and the stalk of streamer, where the variation of the  Mg$^{9+}$ density is small, due to coupling with outflowing protons. The corresponding kinetic temperatures obtained from the model are shown in  Figure~\ref{vrtcr:fig}c. It is evident that the Mg$^{9+}$ ion temperature shows rapid increase reaching 7.5 MK at about 1.6 \rsun, and than decreases up to 3.5 \rsun, matching the proton and electron temperature outside the streamer. The kinetic temperature with error bars of Mg$^{9+}$ ions obtained from observations are shown, and the model results produce good agreement.

\begin{figure}
\center
\includegraphics[width=6.5in,height=3.5in]{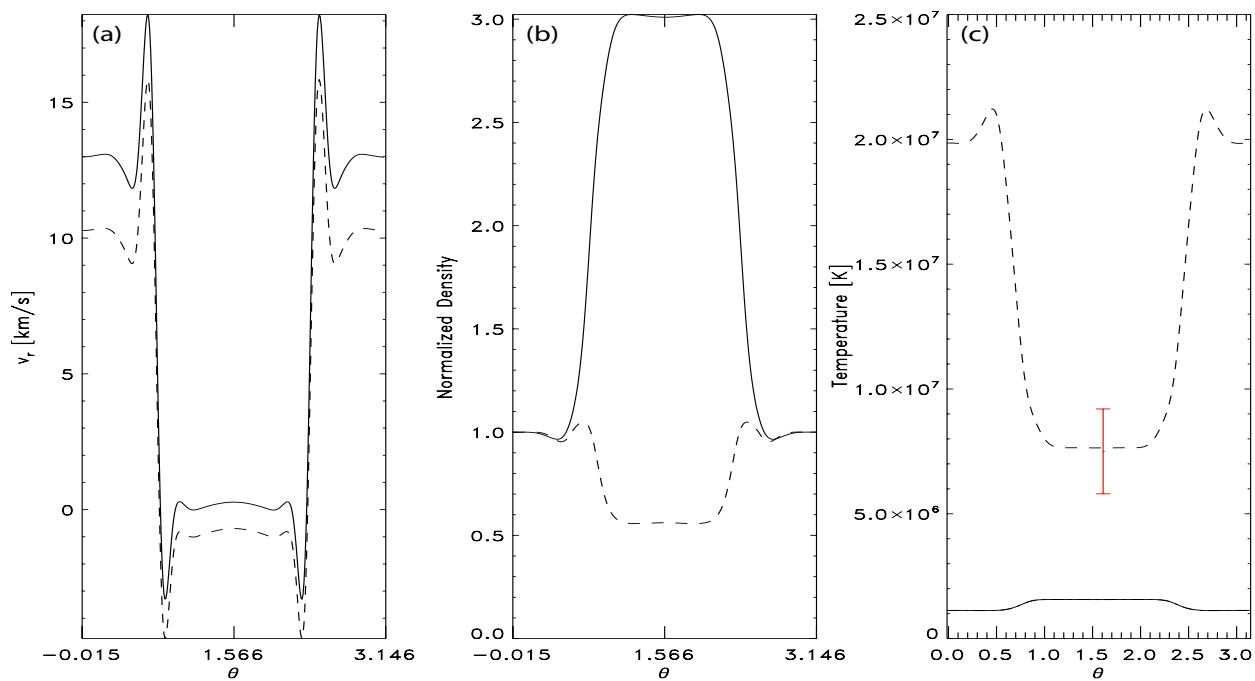}
\caption{The cross-section in the latitudinal ($\theta$) direction  at $r=1.61$ \rsun\, of the streamer calculated with the three-fluid model at $t=98.1\tau_A$ (protons - solid; Mg$^{9+}$ ions - dashes). (a) The radial velocity. (b) The normalized density. (c) The temperature (electrons shown with dotted curve that overlaps the proton temperature). The red symbol with the error bar shows the Mg$^{9+}$ kinetic temperature at $r=1.61$ \rsun\, given in Table~\ref{tk:table}.}
\label{vrtheta:fig}
\end{figure}

The cross-section of the streamer in the latitudinal ($\theta$) direction at $r=1.61$ \rsun\, is shown in Figure~\ref{vrtheta:fig}. The outflow speed is small inside the streamer, consistent with the previous results. The normalized density (to the value at $\theta=0$) of protons and Mg$^{9+}$ ions shows opposite variation inside the closed field region of the streamer compared to the open field region outside. It is evident that the protons density is nearly 3 times larger inside the streamer, compared to the density outside at this height. The Mg$^{9+}$ ion density in the core of the streamer is about half the density outside at this height. The temperature of protons and electrons inside the streamer is increased by about 50\% compared to the outside temperature, while the Mg$^{9+}$ ion temperature is lower by a factor of $\sim2.8$ compared to the temperature outside the closed field region of the streamer at this height, even though the heating function is spherically symmetric. The agreement between the model and the observed Mg$^{9+}$ kinetic temperature in the center of the streamer is evident. The temperature increase outside the streamer is consistent with past observations of emission line broadening of \mgx\ in coronal holes \citep{Koh99}. The nearly steady radial dependence of $T_e$ and $T_p$ in the core of the quiescent streamer is evident.

\begin{figure}
\center
\includegraphics[width=8cm]{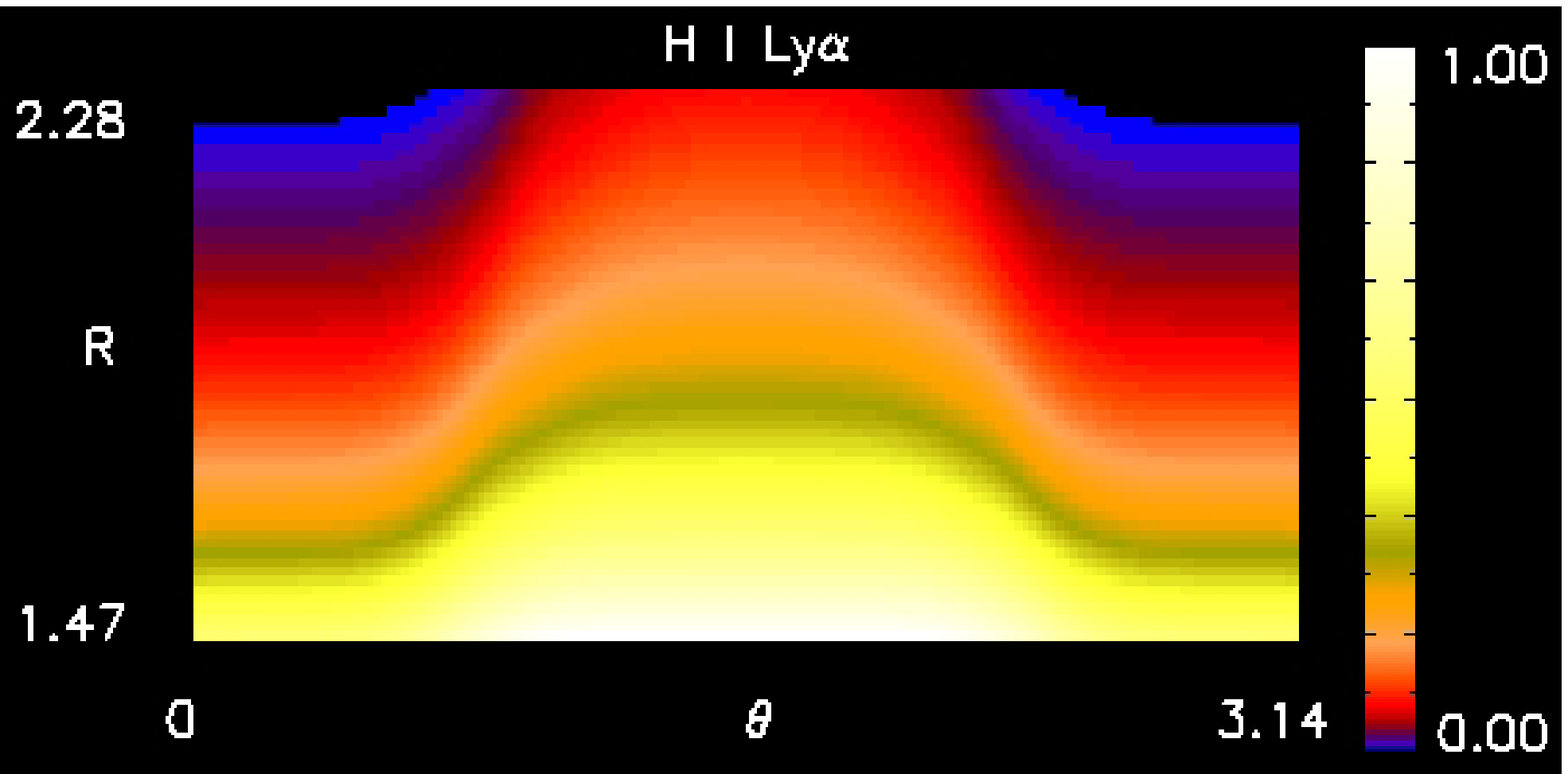}\\
\includegraphics[width=8cm]{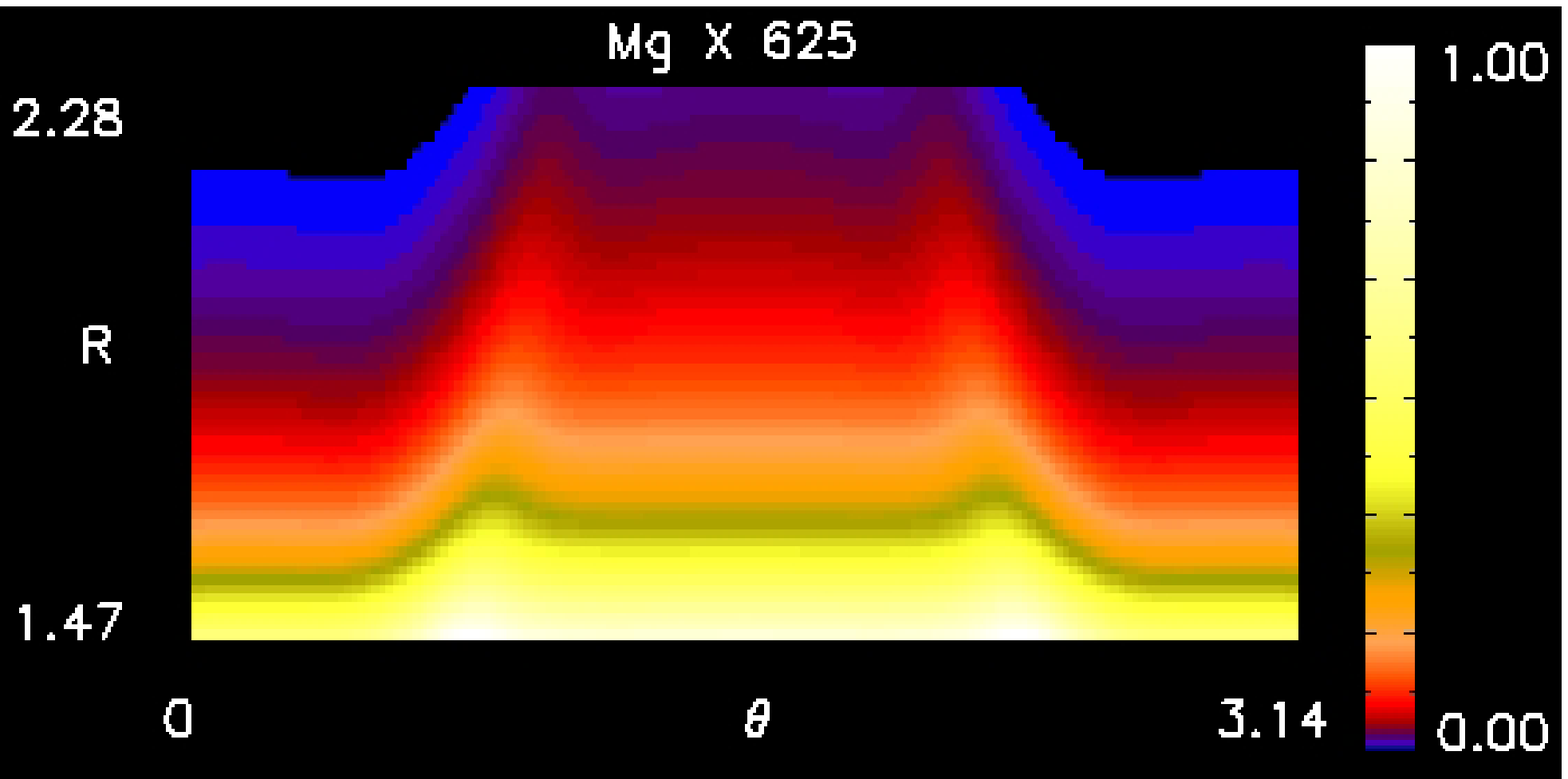}\\
\hspace{-0.9cm}\vspace{0.25cm}\includegraphics[width=7.5cm]{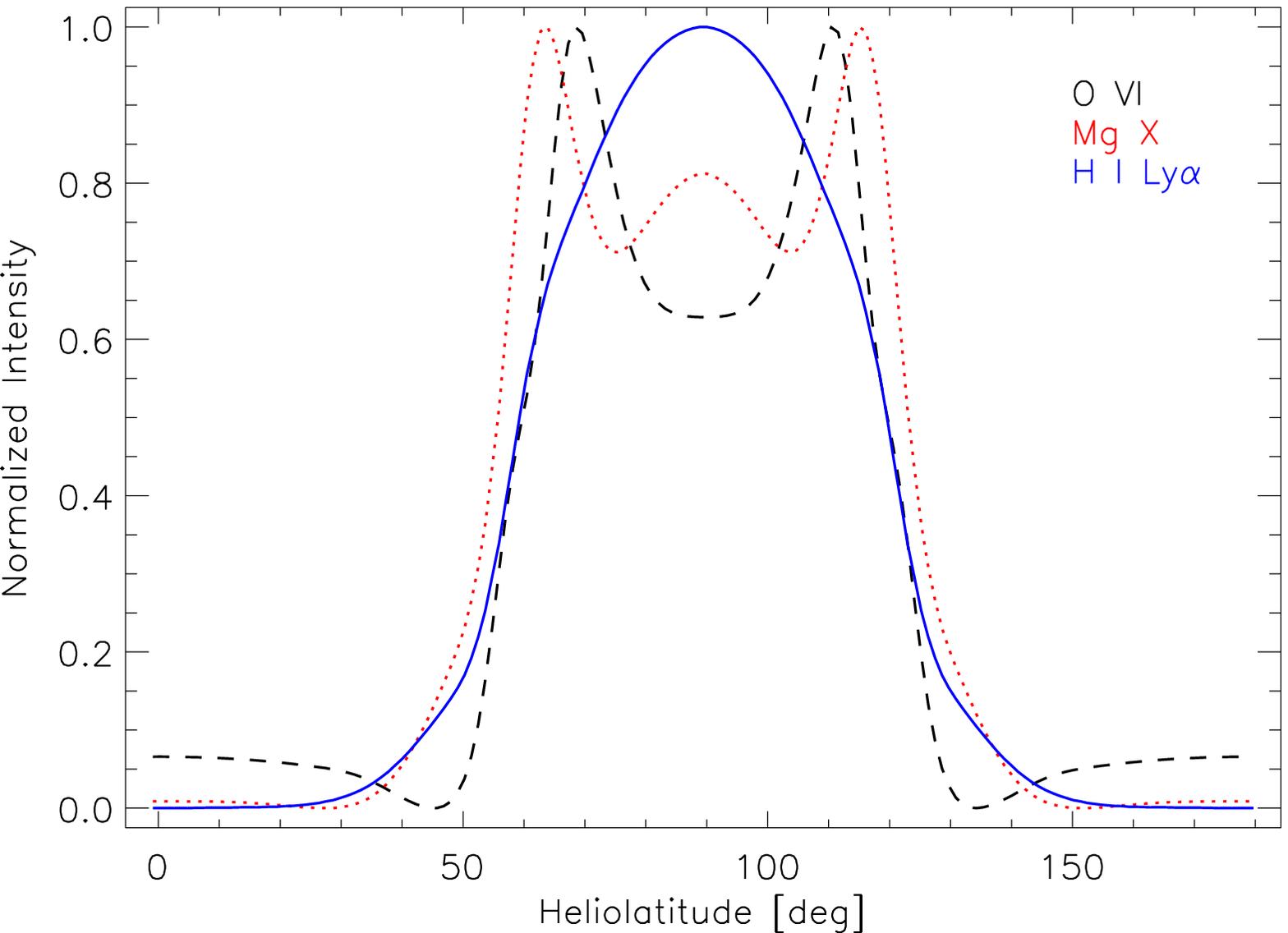}
\caption{The normalized intensity maps of \hi\,\lya\, (top) and of \mgx\, (middle) lines as computed from the model. The latitudinal intensity profiles at 1.63 \rsun\,are shown in the bottom panel (\hi, solid blue line; \mgx, dotted red line, \ovi, dashed black line).}
\label{int_map:fig}
\end{figure}

\begin{figure}
\center
\includegraphics[width=8cm]{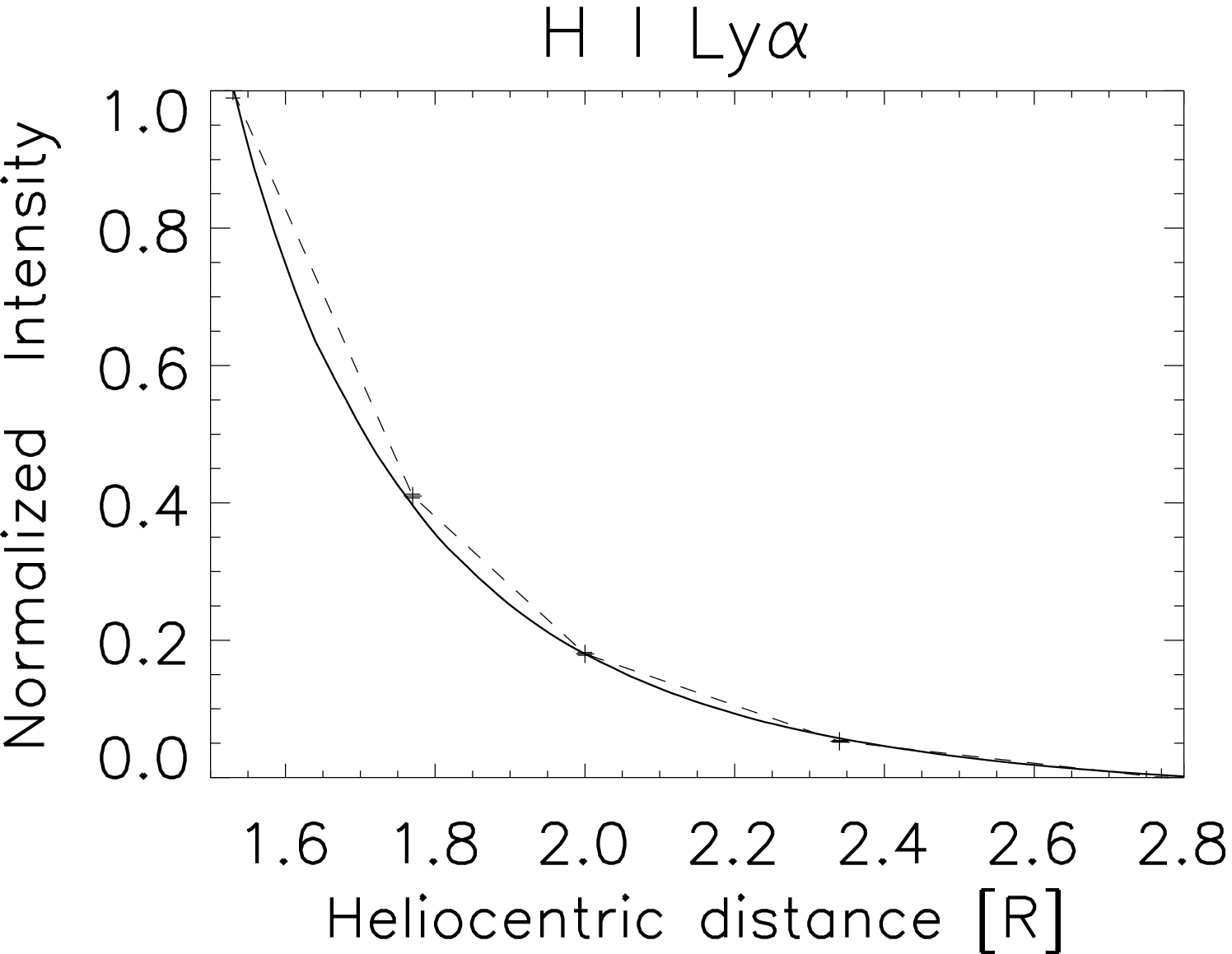}
\includegraphics[width=8cm]{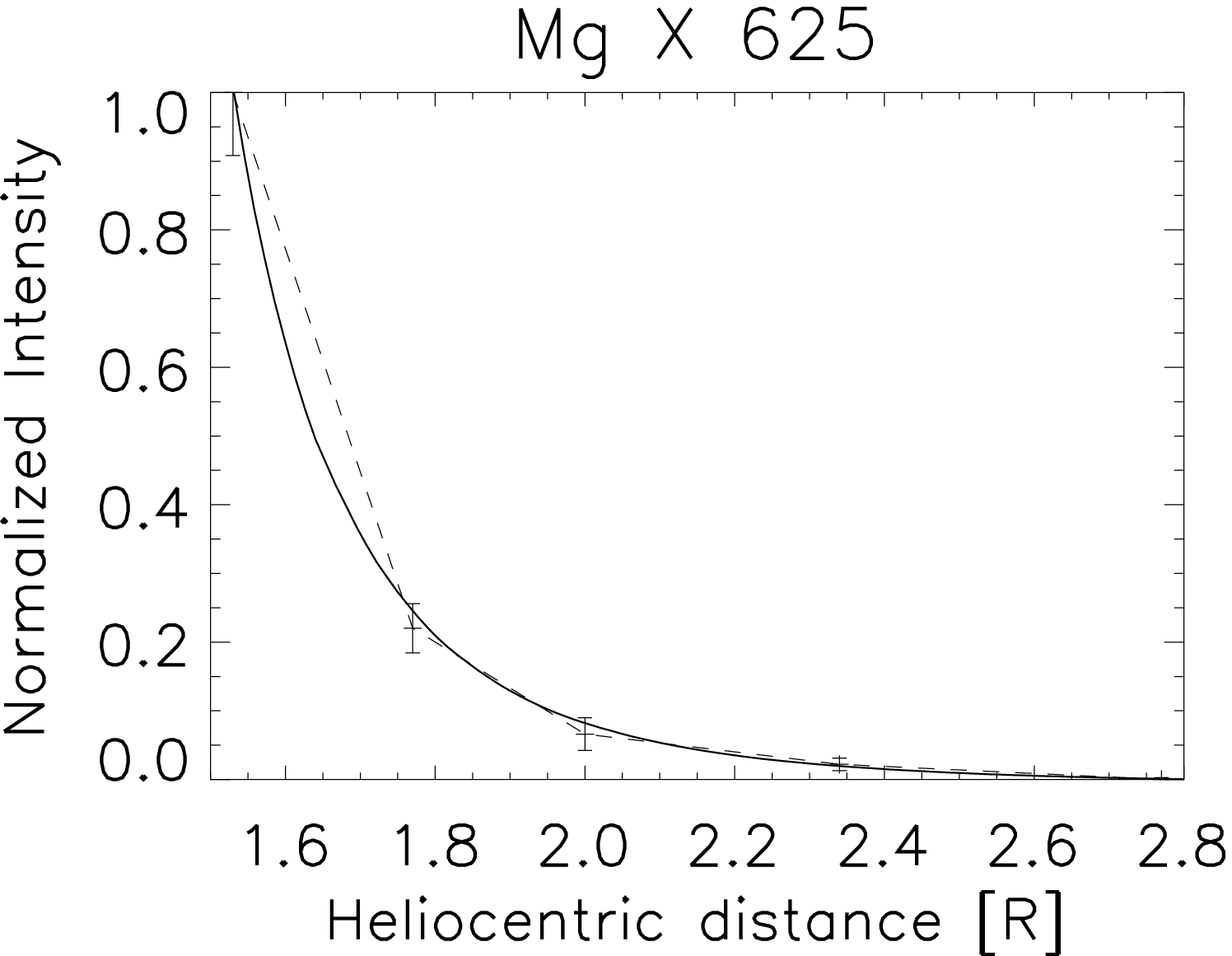}
\caption{The radial intensity profiles (normalized) from 1.5 to 2.8 \rsun\, for \hi\,\lya\, (left) and of \mgx\,(right) lines. The solid  line shows the values computed from the model, the points and the dashed line are the values observed by UVCS.}
\label{int_rad:fig}
\end{figure}

The densities, temperatures, and velocities of the coronal electrons, protons, and Mg$^{9+}$ ions obtained from the 2.5D three-fluid model are used to compute the expected intensity from \hi\,\lya\, and \mgx\, 625\,\AA\ spectral lines, employing the same method as described in Section 4.2 in \citet{OAG11}. In our computation of the radiative component, we assumed uniformly bright exciting radiation coming from the solar disk, in particular for \hi\,\lya\, we adopted the SUMER profile observed at solar minimum in 1996 July \citep{Lem02} and for \mgx\ 625\AA\ the intensity reported by \citet{And12}, and the line width by \citet{Lan90}. For the collisional component, we consider the ionization equilibrium by \citet{Maz98} for \hi\ atoms and by \citet{AR85} for Mg$^{9+}$ ions. The integration along the line-of-sight has been performed by assuming a cylindrical (azimuthal) symmetry and the derived intensity maps are shown in Figure~\ref{int_map:fig} in normalized units, compare well with the images in \lya\, and \mgx\  line intensities shown in Figure~\ref{MgxUV:fig} obtained from the UVCS observations (rotated by 90$^\circ$). In the bottom panel, the latitudinal variation of the normalized intensities are shown for \hi\,\lya\, (solid curve) and for \mgx\,(dotted) lines at 1.63 \rsun, and compare well with the variation shown in the right panel of Figure~\ref{uvcs_strm:fig}. The normalized intensity of \ovi\ computed in \citet{OAG11} is also shown for comparison. Moreover, the radial profiles of \hi\,\lya\, and \mgx\,intensities averaged over the entire streamer region are shown in Figure~\ref{int_rad:fig}. It is evident that the values computed from the model (solid lines) and the values obtained from the UVCS observations on August 23, 1997 (points and dashed lines) are in very good agreement.

\section{Discussion and Conclusions}

The origin of the slow solar wind has been associated with streamers in the past. However, the details of the slow solar wind acceleration process and the outflow regions are poorly known. Past UVCS observations of \ovi\ ion emission and models of quiescent streamers show that the core of the streamer is characterized by lower relative abundance of the O$^{5+}$ ions, and the legs of the streamers show increased relative ion abundance due to collisional interaction (Coulomb friction)  with outflowing protons and electrons. It was suggested in the past   \citep[e.g.][]{Ray97,Noc97,Ofm00a} that O$^{5+}$ ion emission observations can be use as the diagnostic of the slow solar wind acceleration regions in coronal magnetic streamer legs, and the closed magnetic field regions in the quiescent streamer cores due to gravitational settling. Here we show for the first time using observations and detail modeling that these properties hold as well for Mg$^{9+}$ ions and therefore are expected for other heavy ions in quiescent streamers not yet observed spectroscopically in streamers.  

In the present study we analyze UVCS observations of Mg$^{9+}$ EUV emission in the solar minimum streamer, and compare, for the first time, with hydrogen and O$^{5+}$ intensities. Both heavy ion emissions are significantly different in structure from the \hi\,\lya\ emission, showing opposite dependence on latitude. In particular, we find that a streamer core dimming is visible both in \ovi\, and in \mgx\, lines, although it is less prominent in \mgx. We have derived the observed emission line widths at the center of the streamer to calculate the $V_{1/e}$ velocity and the corresponding kinetic temperature, $T_{k,i}$, of the Mg$^{9+}$ ions. We find that in quiescent streamer at solar minimum, the 1/e velocity of Mg$^{9+}$ ions  is comparable or slightly larger than that 
of O$^{5+}$ ions, that is $V_{1/e,MgX} \geq V_{1/e,OVI}$, and that  $T_{k,MgX}\approx 2T_{k,OVI}$ indicating more than mass proportional heavy ion heating.

The three-fluid model developed to study the structure and dynamics of O$^{5+}$ ions has been adapted for the present study of Mg$^{9+}$ ions. The model shows that preferential and more than mass-proportional heating of the heavy ions (compared to protons and electrons) is required in order to reproduce the observed kinetic temperatures of the ions. Interestingly, only radial variation of the empirical ion heating function needs to be specified, while the three-fluid model produces the two-dimensional Mg$^{9+}$ streamer temperature and density structure with good qualitative agreement with the available observations. We found from the three-fluid model that compared to O$^{5+}$ the required heating per particle of Mg$^{9+}$ is significantly larger and must be deposited closer to the Sun to match observations. The higher kinetic temperature of Mg$^{9+}$ ions outside the streamer is consistent with the higher temperature of Mg$^{9+}$ deduced in coronal holes \citep{Koh99}. The model also provides details on the density, temperature and outflow structure of Mg$^{9+}$ ions in the solar minimum streamer not available from observations with present instruments (such as the separation between the core and the legs of the streamer). The output of the model was used to compute the expected \hi\,\lya\, and \mgx\, emission maps and the parameters of the model were iterated to optimized the agreement with observations. Good agreement was found for the radial profiles of the intensities, the core dimming of \mgx\, and the temperature profiles of the ions. We find that the effects of gravitational settling in the quiescent streamer core characterized by close magnetic field, and the Coulomb friction with the outflowing bulk solar wind in open field regions are the dominant processes that affect their relative abundance variation and outflow properties of Mg$^{9+}$ ions.  

\acknowledgments LO would like to acknowledge support by NSF grant AGS-1059838 and NASA grant NNX10AC56G. UVCS
is a joint project of NASA, the Agenzia Spaziale Italiana (ASI),
and the Swiss Founding Agencies. LA has been funded through contract I/023/09/0 and I/013/12/0 between the National Institute for Astrophysics (INAF) and the Agenzia Spaziale Italiana (ASI).


\begin{thebibliography}{33}
\expandafter\ifx\csname natexlab\endcsname\relax\def\natexlab#1{#1}\fi

\bibitem[{{Abbo} {et~al.}(2010{\natexlab{a}}){Abbo}, {Antonucci}, {Miki{\'c}},
  {Linker}, {Riley}, \& {Lionello}}]{Abb10b}
{Abbo}, L., {Antonucci}, E., {Miki{\'c}}, Z., {Linker}, J.~A., {Riley}, P., \&
  {Lionello}, R. 2010{\natexlab{a}}, Advances in Space Research, 46, 1400

\bibitem[{{Abbo} {et~al.}(2010{\natexlab{b}}){Abbo}, {Ofman}, \&
  {Giordano}}]{Abb10a}
{Abbo}, L., {Ofman}, L., \& {Giordano}, S. 2010{\natexlab{b}}, Twelfth
  International Solar Wind Conference, ed. M. Maksimovic et al. (Melville, NY: AIP), 1216, 387

\bibitem[{{Akinari}(2007)}]{Aki07}
{Akinari}, N. 2007, \apj, 668, 1196

\bibitem[{{Andretta} {et~al.}(2012){Andretta}, {Telloni}, \& {Del
  Zanna}}]{And12}
{Andretta}, V., {Telloni}, D., \& {Del Zanna}, G. 2012, \solphys, 279, 53

\bibitem[{{Antonucci} {et~al.}(2005){Antonucci}, {Abbo}, \& {Dodero}}]{Ant05}
{Antonucci}, E., {Abbo}, L., \& {Dodero}, M.~A. 2005, \aap, 435, 699

\bibitem[{{Arnaud} \& {Rothenflug}(1985)}]{AR85}
{Arnaud}, M., \& {Rothenflug}, R. 1985, \aaps, 60, 425

\bibitem[{{Chen} \& {Li}(2004)}]{CL04}
{Chen}, Y., \& {Li}, X. 2004, \apjl, 609, L41

\bibitem[{{Cranmer} {et~al.}(1999){Cranmer}, {Field}, \& {Kohl}}]{CFK99}
{Cranmer}, S.~R., {Field}, G.~B., \& {Kohl}, J.~L. 1999, \apj, 518, 937

\bibitem[{{Esser} {et~al.}(1999){Esser}, {Fineschi}, {Dobrzycka}, {Habbal},
  {Edgar}, {Raymond}, {Kohl}, \& {Guhathakurta}}]{Ess99}
{Esser}, R., {Fineschi}, S., {Dobrzycka}, D., {Habbal}, S.~R., {Edgar}, R.~J.,
  {Raymond}, J.~C., {Kohl}, J.~L., \& {Guhathakurta}, M. 1999, \apjl, 510, L63

\bibitem[{{Kohl} {et~al.}(1999){Kohl}, {Fineschi}, {Esser}, {Ciaravella},
  {Cranmer}, {Gardner}, {Suleiman}, {Noci}, \& {Modigliani}}]{Koh99}
{Kohl}, J.~L., {Fineschi}, S., {Esser}, R., {Ciaravella}, A., {Cranmer}, S.~R.,
  {Gardner}, L.~D., {Suleiman}, R., {Noci}, G., \& {Modigliani}, A. 1999, Space
  Science Reviews, 87, 233

\bibitem[{{Kohl} {et~al.}(1997){Kohl}, {Noci}, {Antonucci}, {Tondello},
  {Huber}, {Gardner}, {Nicolosi}, {Strachan}, {Fineschi}, {Raymond}, {Romoli},
  {Spadaro}, {Panasyuk}, {Siegmund}, {Benna}, {Ciaravella}, {Cranmer},
  {Giordano}, {Karovska}, {Martin}, {Michels}, {Modigliani}, {Naletto},
  {Pernechele}, {Poletto}, \& {Smith}}]{Koh97}
{Kohl}, J.~L., {Noci}, G., {Antonucci}, E., {Tondello}, G., {Huber}, M.~C.~E.,
  {Gardner}, L.~D., {Nicolosi}, P., {Strachan}, L., {Fineschi}, S., {Raymond},
  J.~C., {Romoli}, M., {Spadaro}, D., {Panasyuk}, A., {Siegmund}, O.~H.~W.,
  {Benna}, C., {Ciaravella}, A., {Cranmer}, S.~R., {Giordano}, S., {Karovska},
  M., {Martin}, R., {Michels}, J., {Modigliani}, A., {Naletto}, G.,
  {Pernechele}, C., {Poletto}, G., \& {Smith}, P.~L. 1997, \solphys, 175, 613

\bibitem[{{Landini} \& {Monsignori Fossi}(1990)}]{Lan90}
{Landini}, M., \& {Monsignori Fossi}, B.~C. 1990, \aaps, 82, 229

\bibitem[{{Lemaire} {et~al.}(2002){Lemaire}, {Emerich}, {Vial}, {Curdt},
  {Sch{\"u}hle}, \& {Wilhelm}}]{Lem02}
{Lemaire}, P., {Emerich}, C., {Vial}, J., {Curdt}, W., {Sch{\"u}hle}, U., \&
  {Wilhelm}, K. 2002, in ESA Special Publication, Vol. 508, From Solar Min to
  Max: Half a Solar Cycle with SOHO, ed. {A.~Wilson} (Noordwijk: ESA), 219--222

\bibitem[{{Lenz}(2004)}]{Len04}
{Lenz}, D.~D. 2004, \apj, 604, 433

\bibitem[{{Marocchi} {et~al.}(2001){Marocchi}, {Antonucci}, \&
  {Giordano}}]{Mar01}
{Marocchi}, D., {Antonucci}, E., \& {Giordano}, S. 2001, Annales Geophysicae,
  19, 135

\bibitem[{{Mazzotta} {et~al.}(1998){Mazzotta}, {Mazzitelli}, {Colafrancesco},
  \& {Vittorio}}]{Maz98}
{Mazzotta}, P., {Mazzitelli}, G., {Colafrancesco}, S., \& {Vittorio}, N. 1998,
  \aaps, 133, 403

\bibitem[{{Morgan} {et~al.}(2008){Morgan}, {Fineschi}, {Habbal}, \&
  {Li}}]{Mor08}
{Morgan}, H., {Fineschi}, S., {Habbal}, S.~R., \& {Li}, B. 2008, \aap, 482, 981

\bibitem[{{Noci} {et~al.}(1997){Noci}, {Kohl}, {Antonucci}, {Tondello},
  {Huber}, {Fineschi}, {Gardner}, {Naletto}, {Nicolosi}, {Raymond}, {Romoli},
  {Spadaro}, {Siegmund}, {Benna}, {Ciaravella}, {Giordano}, {Michels},
  {Modigliani}, {Panasyuk}, {Pernechele}, {Poletto}, {Smith}, \&
  {Strachan}}]{Noc97}
{Noci}, G., {Kohl}, J.~L., {Antonucci}, E., {Tondello}, G., {Huber}, M.~C.~E.,
  {Fineschi}, S., {Gardner}, L.~D., {Naletto}, G., {Nicolosi}, P., {Raymond},
  J.~C., {Romoli}, M., {Spadaro}, D., {Siegmund}, O.~H.~W., {Benna}, C.,
  {Ciaravella}, A., {Giordano}, S., {Michels}, J., {Modigliani}, A.,
  {Panasyuk}, A., {Pernechele}, C., {Poletto}, G., {Smith}, P.~L., \&
  {Strachan}, L. 1997, Advances in Space Research, 20, 2219

\bibitem[{{Ofman}(2000)}]{Ofm00a}
{Ofman}, L. 2000, \grl, 27, 2885

\bibitem[{{Ofman}(2004)}]{Ofm04a}
---. 2004, Advances in Space Research, 33, 681

\bibitem[{{Ofman} {et~al.}(2011){Ofman}, {Abbo}, \& {Giordano}}]{OAG11}
{Ofman}, L., {Abbo}, L., \& {Giordano}, S. 2011, \apj, 734, 30

\bibitem[{{Ofman} \& {Kramar}(2010)}]{OK10}
{Ofman}, L., \& {Kramar}, M. 2010, in Astronomical Society of the Pacific
  Conference Series, Vol. 428, Astronomical Society of the Pacific Conference
  Series, ed. {S.~R.~Cranmer, J.~T.~Hoeksema, \& J.~L.~Kohl}, 321--324

\bibitem[{{Pneuman} \& {Kopp}(1971)}]{PK71}
{Pneuman}, G.~W., \& {Kopp}, R.~A. 1971, \solphys, 18, 258

\bibitem[{{Raymond} {et~al.}(1997){Raymond}, {Kohl}, {Noci}, {Antonucci},
  {Tondello}, {Huber}, {Gardner}, {Nicolosi}, {Fineschi}, {Romoli}, {Spadaro},
  {Siegmund}, {Benna}, {Ciaravella}, {Cranmer}, {Giordano}, {Karovska},
  {Martin}, {Michels}, {Modigliani}, {Naletto}, {Panasyuk}, {Pernechele},
  {Poletto}, {Smith}, {Suleiman}, \& {Strachan}}]{Ray97}
{Raymond}, J.~C., {Kohl}, J.~L., {Noci}, G., {Antonucci}, E., {Tondello}, G.,
  {Huber}, M.~C.~E., {Gardner}, L.~D., {Nicolosi}, P., {Fineschi}, S.,
  {Romoli}, M., {Spadaro}, D., {Siegmund}, O.~H.~W., {Benna}, C., {Ciaravella},
  A., {Cranmer}, S., {Giordano}, S., {Karovska}, M., {Martin}, R., {Michels},
  J., {Modigliani}, A., {Naletto}, G., {Panasyuk}, A., {Pernechele}, C.,
  {Poletto}, G., {Smith}, P.~L., {Suleiman}, R.~M., \& {Strachan}, L. 1997,
  \solphys, 175, 645

\bibitem[{{Steinolfson} \& {Nakagawa}(1976)}]{SN76}
{Steinolfson}, R.~S., \& {Nakagawa}, Y. 1976, \apj, 207, 300

\bibitem[{{Strachan} {et~al.}(2002){Strachan}, {Suleiman}, {Panasyuk},
  {Biesecker}, \& {Kohl}}]{Str02}
{Strachan}, L., {Suleiman}, R., {Panasyuk}, A.~V., {Biesecker}, D.~A., \&
  {Kohl}, J.~L. 2002, \apj, 571, 1008

\bibitem[{{Uzzo} {et~al.}(2003){Uzzo}, {Ko}, {Raymond}, {Wurz}, \&
  {Ipavich}}]{Uzz03}
{Uzzo}, M., {Ko}, Y., {Raymond}, J.~C., {Wurz}, P., \& {Ipavich}, F.~M. 2003,
  \apj, 585, 1062

\bibitem[{{Uzzo} {et~al.}(2004){Uzzo}, {Ko}, \& {Raymond}}]{Uzz04}
{Uzzo}, M., {Ko}, Y.-K., \& {Raymond}, J.~C. 2004, \apj, 603, 760

\bibitem[{{Uzzo} {et~al.}(2007){Uzzo}, {Strachan}, \& {Vourlidas}}]{Uzz07}
{Uzzo}, M., {Strachan}, L., \& {Vourlidas}, A. 2007, \apj, 671, 912

\bibitem[{{Uzzo} {et~al.}(2006){Uzzo}, {Strachan}, {Vourlidas}, {Ko}, \&
  {Raymond}}]{Uzz06}
{Uzzo}, M., {Strachan}, L., {Vourlidas}, A., {Ko}, Y.-K., \& {Raymond}, J.~C.
  2006, \apj, 645, 720

\bibitem[{{V{\'a}squez} \& {Raymond}(2005)}]{Vas05}
{V{\'a}squez}, A.~M., \& {Raymond}, J.~C. 2005, \apj, 619, 1132

\bibitem[{{Withbroe} {et~al.}(1982){Withbroe}, {Kohl}, {Weiser}, \&
  {Munro}}]{Wit82}
{Withbroe}, G.~L., {Kohl}, J.~L., {Weiser}, H., \& {Munro}, R.~H. 1982, \ssr,
  33, 17

\bibitem[{{Zangrilli} {et~al.}(2001){Zangrilli}, {Poletto}, {Biesecker}, \&
  {Raymond}}]{Zan01}
{Zangrilli}, L., {Poletto}, G., {Biesecker}, D., \& {Raymond}, J.~C. 2001, in
  American Institute of Physics Conference Series, Vol. 598, Joint SOHO/ACE
  workshop ''Solar and Galactic Composition'', ed. R.~F.
  {Wimmer-Schweingruber}, 71--76

\end{thebibliography}

\clearpage
\begin{table}[h!]
	\centering
	\caption{The most probable velocity V$_{1/e}$, and the corresponding kinetic temperatures $T_{k,i}$ of \mgx\, and \ovi\, lines observed in quiescent streamers by UVCS.}
	\vspace{0.2cm}
	\begin{tabular}{|l|c|c|c|c|c|c|}
		\hline
Height	&	Date		&	Slit	&	$V_{1/e, MgX}$  & $T_{k, MgX}$  & $V_{1/e, OVI}$ & $T_{k, OVI}$ \\
\rsun	&YY.MM.DD.hh&	$\mu m$&	km/s	&   MK	& km/s  &  MK \\\hline
1.39 	&	96.08.16.17	&	31	&	60 $\pm$13	& 5.2 $\pm$2.3 &	50 $\pm$ 5 & 2.4 $\pm$0.5 \\
1.49	&	96.07.15.18	&	76	&	71 $\pm$7 & 7.3 $\pm$1.4  &	70 $\pm$ 3 & 4.8 $\pm$0.4\\
1.61	&	97.07.31.19	&	100	&	72 $\pm$8 & 7.5 $\pm$ 1.7	&	58 $\pm$ 3 & 3.3 $\pm$0.3 \\
1.70 	&	96.07.23.16	&	97	&	70 $\pm$7 & 7.1 $\pm$ 1.4	&	63 $\pm$ 3 & 3.9 $\pm$0.4 \\
		\hline
	\end{tabular}
	\label{tk:table}
\end{table}

\end{document}